\documentclass[final,15pt,5p]{elsarticle}

\usepackage{lineno,hyperref,amsmath}
\usepackage{color}
\usepackage{here}
\journal{Journal of \LaTeX\ Templates}

\makeatletter
  \AtBeginDocument{%
    \def\include@graphics#1{%
      \begingroup\fboxsep=-\fboxrule
      \fbox{\rule{\@ifundefined{Gin@@ewidth}{150pt}{\Gin@@ewidth}}{0pt}%
        \rule{0pt}{\@ifundefined{Gin@@eheight}{200pt}{\Gin@@eheight}}}\endgroup}}
        \makeatother

\begin{document}

\begin{frontmatter}

\title{Performance of the PRAXyS X-ray Polarimeter}

\author[RIKEN]{W.B.~Iwakiri}\corref{correspond}

\cortext[correspond]{Corresponding author. Tel:+81-48-462-4874; FAX:+81-48-462-4640}
\ead{wataru.iwakiri@riken.jp}
\author[Goddard,Rock]{J.K.~Black}
\author[Goddard]{R.~Cole}
\author[Hakubi,Kyoto]{T.~Enoto}
\author[RIKEN]{A.~Hayato}
\author[Goddard]{J.E.~Hill}
\author[Goddard]{K.~Jahoda}
\author[Iowa]{P.~Kaaret}
\author[Hiroshima]{T.~Kitaguchi}
\author[Rika,RIKEN]{M.~Kubota}
\author[Iowa]{H.~Marlowe}
\author[Iowa]{R.~McCurdy}
\author[Rika,RIKEN]{Y.~Takeuchi}
\author[RIKEN,Rika]{T.~Tamagawa}

\address[RIKEN]{RIKEN Nishina Center, 2-1 Hirosawa, Wako, Saitama 351-0198, Japan}
\address[Goddard]{NASA Goddard Space Flight Center, Greenbelt, MD 20771, USA}
\address[Rock]{Rock Creek Scientific, 1400 East-West Hwy, Silver Spring, MD, 20910, USA}
\address[Hakubi]{The Hakubi Center for Advanced Research, Kyoto University, Kyoto 606-8302, Japan}
\address[Kyoto]{Department of Astronomy, Kyoto University, Kitashirakawa-Oiwake-cho, Sakyo-ku, Kyoto 606-8502, Japan}
\address[Hiroshima]{Department of Physical Science, Hiroshima University, 1-3-1 Kagamiyama, Higashi-Hiroshima, Hiroshima 739-8526, Japan}
\address[Iowa]{University of Iowa, Iowa City, IA, 52242, USA}
\address[Rika]{Department of Physics, Tokyo University of Science, 3-1 Kagurazaka, Shinjuku-ku, Tokyo 162-8601, Japan}

\begin{abstract}
The performance of the Time Projection Chamber (TPC) polarimeter for the Polarimeter for Relativistic Astrophysical X-ray Sources (PRAXyS)
Small Explorer was evaluated using polarized and unpolarized X-ray sources.  The PRAXyS mission will enable exploration of the universe
through X-ray polarimetry in the 2--10 keV energy band. We carried out performance tests of the polarimeter at the Brookhaven National
Laboratory, National Synchrotron Light Source (BNL-NSLS) and at NASA's Goddard Space Flight Center. The polarimeter was tested with linearly polarized,
monochromatic X-rays at 11 different energies between 2.5 and 8.0 keV. At maximum sensitivity, the measured modulation factors at 2.7, 4.5 and 8.0 keV are 27\%, 43\% and 59\%,
respectively and the measured angle of polarization is consistent with the expected value at all energies. Measurements with a broadband, unpolarized X-ray source placed
a limit of less than 1\% on false polarization in the PRAXyS polarimeter.

\end{abstract}


\end{frontmatter}

%\linenumbers

\section{Introduction}
Cosmic X-ray polarimetry is a powerful technique for
studying the physics of extreme environments such as strong
gravitational fields and magnetic fields in the universe. For example,
it will be possible to observe vacuum polarization effects in the
extreme magnetic fields of magnetized neutron stars, where the fields
are 10$^{12}$ G or greater \cite{meszaros1985,ghosh2013}.  
However,  X-ray polarization measurements below 10 keV have only succeeded for the Crab Nebula with measurements made by a Bragg scattering polarimeter on a sounding rocket and on the {\it{OSO-8}} satellite in the
1970s \cite{novick1972,weisskopf1978a}.
In the interim, photoelectric polarimeters with greater sensitivity have been developed, first using CCDs \cite{tsunemi1992, buschhorn1994} and more recently gas detectors \cite{costa2001,kevin2007}. 

To maximize sensitivity we have developed a gas polarimeter that employs the Time Projection Chamber (TPC) technique \cite{kevin2007,kevin2010,joe2012,joe2014,enoto2014}. In this case the detection plane is parallel to the incident X-rays. This design allows the detector depth (and efficiency) to be increased without also increasing diffusion (limiting sensitivity). This advantage comes at the expense of true imaging of the sky. However, black holes and neutron stars have angular scales well below micro-arcsecond and sky imaging is of limited scientific utility.

The Polarimeter for Relativistic Astrophysical X-ray Sources (PRAXyS), based on this TPC polarimeter, has been selected
for Phase A study as one of three Small Explorer (SMEX)
missions. PRAXyS is designed to make highly sensitive measurements of the linear X-ray
polarization of astronomical sources in the 2--10 keV
energy band. The primary observational goals of PRAXyS are to observe
a sample of black holes and neutron stars 
brighter than $2 \times 10^{-11}$ ergs s$^{-1}$ cm$^{-2}$ in the 2--10 keV band, with a sensitivity to polarization fractions as small as 1\%. This paper
reports the polarization sensitivity of the TPC polarimeter and upper
limits to the systematic errors.

Photoelectric polarimeters exploit the intrinsic polarization sensitivity of photoelectric absorption.   The photoelectron
produced by the interaction of an X-ray with a gas atom creates an ionization track. The initial direction of the ionization
track contains information about the X-ray polarization. Gas detectors use an electric field to drift the ionization
track to a multiplication and detection region. Costa et al. \cite{costa2001} first proposed a design in which the drift field is
parallel to the X-ray direction of incidence.  For this concept, maximizing sensitivity requires balancing the greater detection efficiency afforded by deeper detectors with the degraded sensitivity caused by the increased diffusion as tracks drift greater distances. The TPC polarimeter breaks this competition by drifting the track perpendicular to the incident 
direction. This allows greater efficiency albeit at the cost of using two different detector properties to create a two
dimensional image of the track \cite{kevin2007}. 

PRAXyS employs a TPC polarimeter in which the charge detection plane consists
of a Gas Electron Multiplier (GEM) designed by RIKEN \cite{tamagawa2009} mounted over strip anodes parallel to the
incident X-rays. Two-dimensional images of photoelectron tracks are created using a one-dimensional strip readout and by timing the
arrival of charge \cite{kevin2007}. The readout and detector plane is described by Hill et al. 2014 \cite{joe2014}. An estimate of the initial track direction is obtained from each event image.

Combining the quantum mechanical expectations for
K-shell absorption and instrumental imperfections, one expects a
measured distribution of photoelectron emission angles:
\begin{equation}
N(\phi) = A + B\text{cos}^2(\phi - \phi_0), 
\label{eq1}
\end{equation}
where $\phi$ represents the azimuthal angle of the photoelectron
track, $\phi_0$ is the source polarization angle, and the constants
$A$ and $B$ are characteristics of the detector and are typically
dependent on energy \cite{costa2001,tod2013}. A histogram of reconstructed
emission angles is called a ``modulation curve''. The amplitude of a modulation curve $a$ is defined as,

\begin{equation}
a = \frac{f_{\text{max}} - f_{\text{min}}}{f_{\text{max}} + f_{\text{min}}} = \frac{B}{2A+B},
\label{eq2}
\end{equation}
where $f_{\text{max}}$ and $f_{\text{min}}$ are the maximum and minimum value of the
modulation curve.  The analyzing power of a polarimeter, called the modulation factor, $\mu$, is the amplitude for 100\% polarized input. The polarization fraction, $a_p$, of a source is then given by $a_p = a / \mu$.

%\begin{equation}
%a_p = \biggl(\frac{a}{\mu}\biggr).
%\label{eq3}
%\end{equation}

To measure the polarization of astrophysical sources we must know the polarimeter's response to both 100\% polarized and unpolarized X-rays across the energy
band. We present experimental results for nearly 100\% polarized X-rays in section 3.4 and unpolarized X-rays in section 3.5. 

The smallest polarization which would not be observed by chance, with 99\% confidence, is inversely proportional to $\mu$ and inversely proportional to the square root of the number of photons. This Minimum Detectable Polarization (MDP) is given by:

\begin{equation}
\text{MDP} = \frac{4.29}{\mu \sqrt{F_\text{s} A_{\text{eff}}{\epsilon}T}} \sqrt{1 + \frac{R_\text{b}}{F_\text{s} A_{\text{eff}}{\epsilon}} },
\end{equation}
where $F_\text{s}$ is the source flux, $A_{\text{eff}}$ the mirror effective area, $\epsilon$ the polarimeter quantum efficiency, $R_\text{b}$ the background count rate,
and $T$ is the observation time. The coefficient, $4.29$ corresponds to 99\% confidence that the signal is not created by chance \cite{weisskopf2009}. 
In terms of detector parameters, the MDP scales as $1/\mu\sqrt{\epsilon}$ assuming the background count rate is negligible. The figure-of-merit for the polarimeter, which minimizes the MDP, is then $\mu \sqrt{\epsilon}$, where $\epsilon$ includes losses due to, for example, an X-ray window and events rejected in analysis. To achieve a statistical precision estimated by the MDP formula, systematic errors must be lower than the MDP. 

\section{Experimental Setup}

We measured the polarization sensitivity of the PRAXyS polarimeter using a single detector module of the flight design. The energy dependent sensitivity to $\sim$ 100\% polarized X-rays over the full range of detector interaction positions was measured at the X-19A beamline at Brookhaven National Laboratory, National Synchrotron Light Source (BNL-NSLS) facility in September 2014. To search for systematic errors that would create a false modulation, we measured the response to a broadband, unpolarized source with a bremsstrahlung spectrum, at the NASA Goddard Space Flight Center (NASA/GSFC).

\begin{figure}[!t]
\begin{center}
\includegraphics[clip,width=7.0cm]{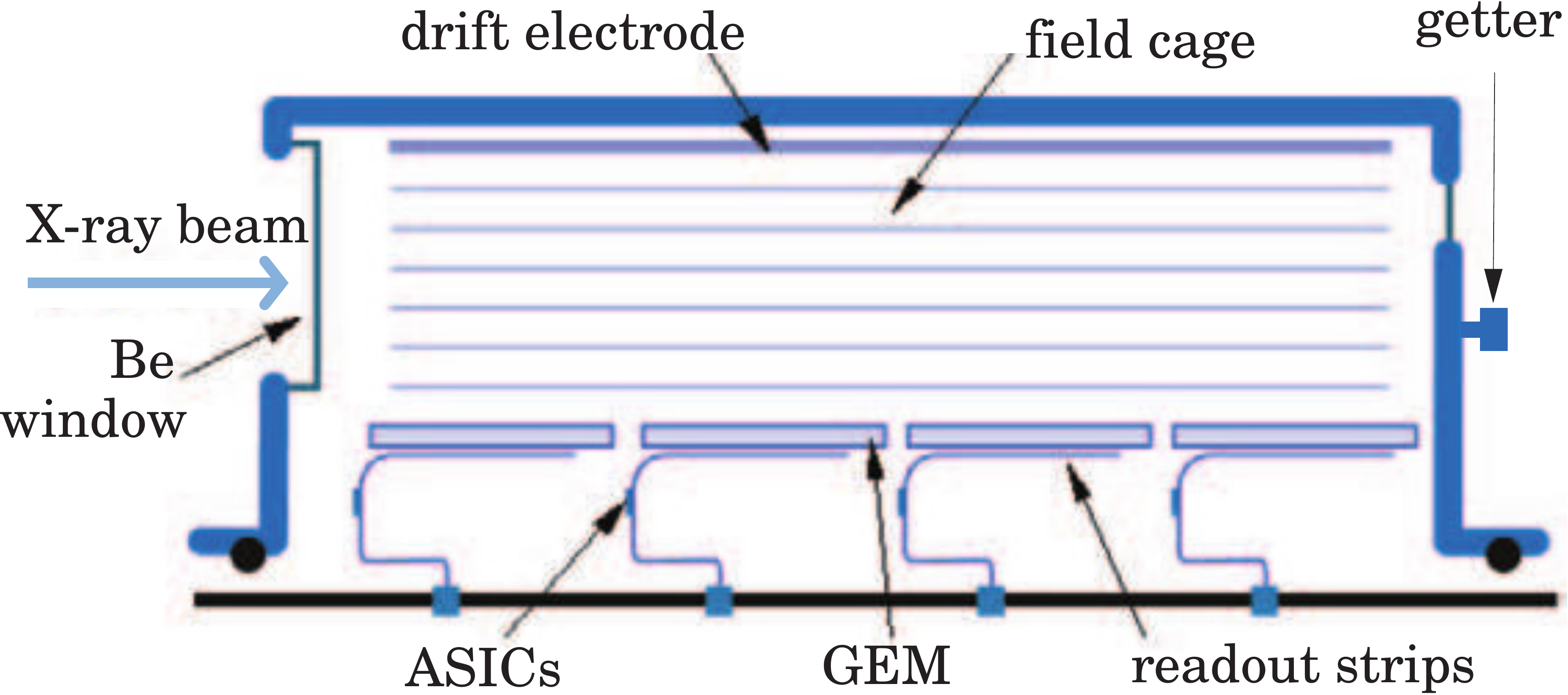}
\caption{Cross-section of the PRAXyS flight polarimeter.}
\label{fig1}
\end{center}
\end{figure}

\begin{figure}[!t]
\begin{center}
\includegraphics[clip,width=8.0cm]{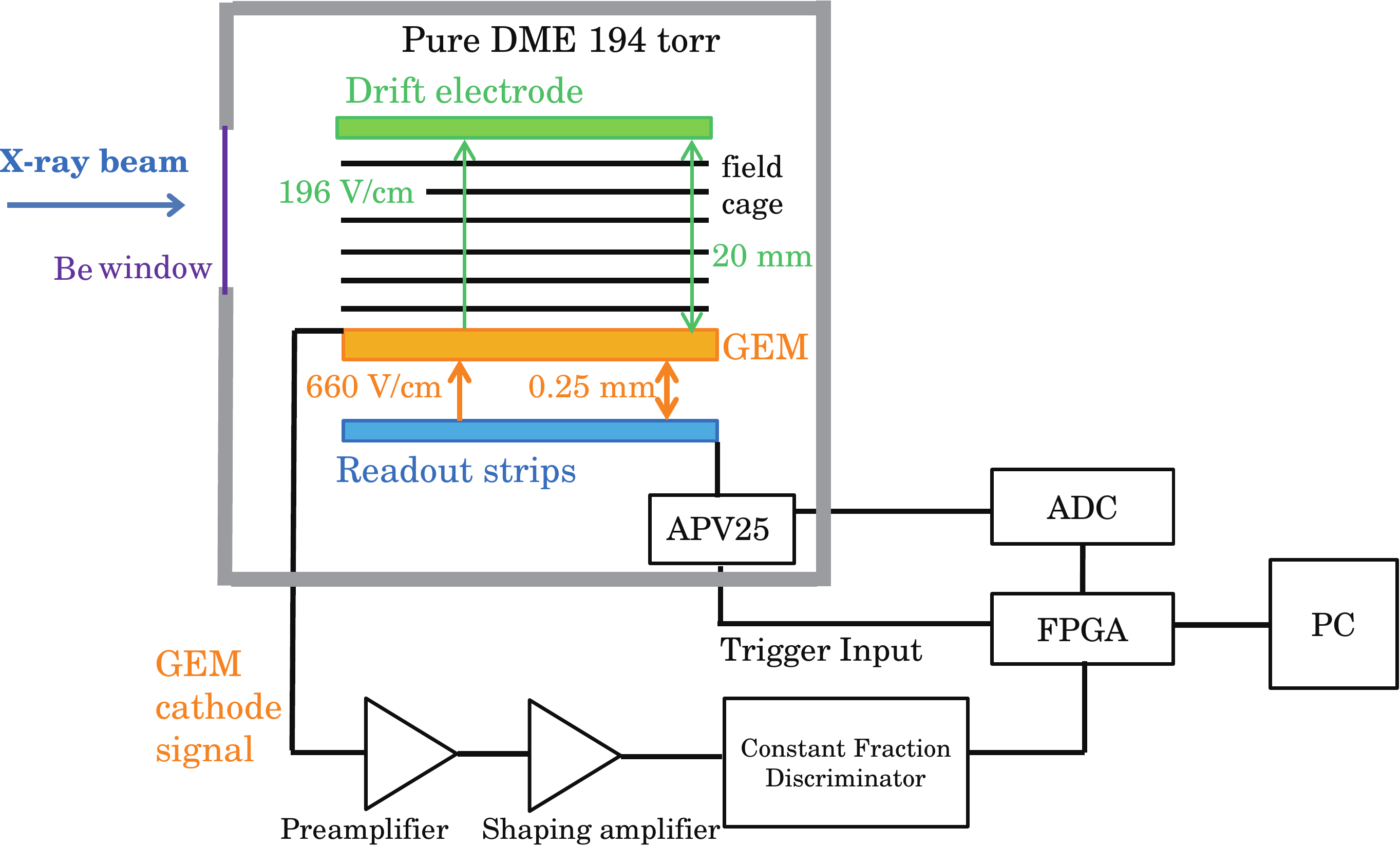}
\caption{Test setup for performance tests of the prototype polarimeter module for PRAXyS using the unpolarized X-ray source and the linear polarized X-ray beam. The electronics inside the gas volume employs the flight design. GSE electronics are employed outside.}
\label{fig2}
\end{center}
\end{figure}

\begin{figure}[!h]
\begin{center}
\includegraphics[clip,width=6.0cm]{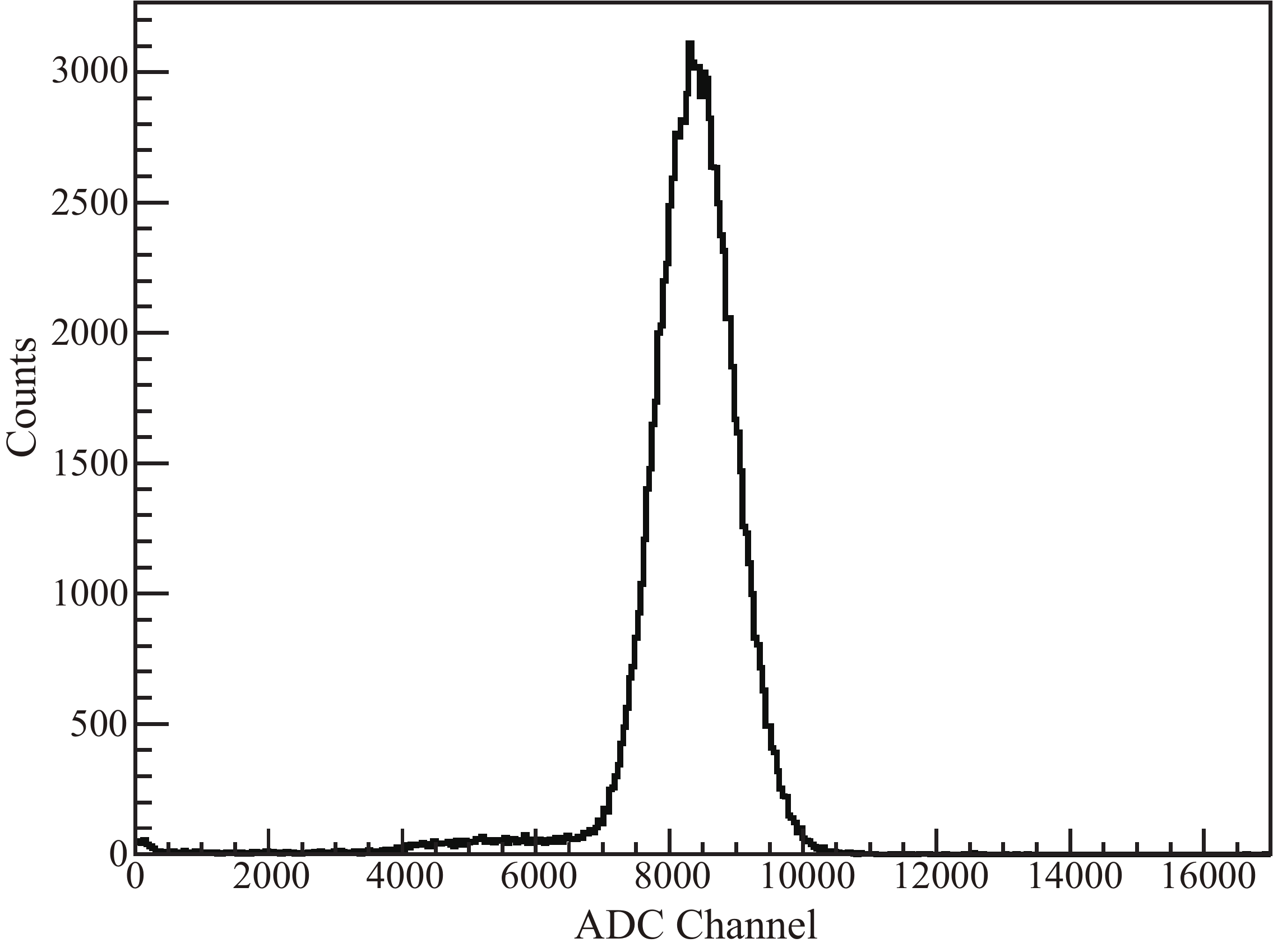}
\caption{Typical ADC spectrum of 6.4 keV X-rays. The energy resolution (FWHM) is about 16\% at 6.4 keV.}
\label{fig3}
\end{center}
\end{figure}

\begin{figure}[!h]
\begin{center}
\includegraphics[clip,width=7.0cm]{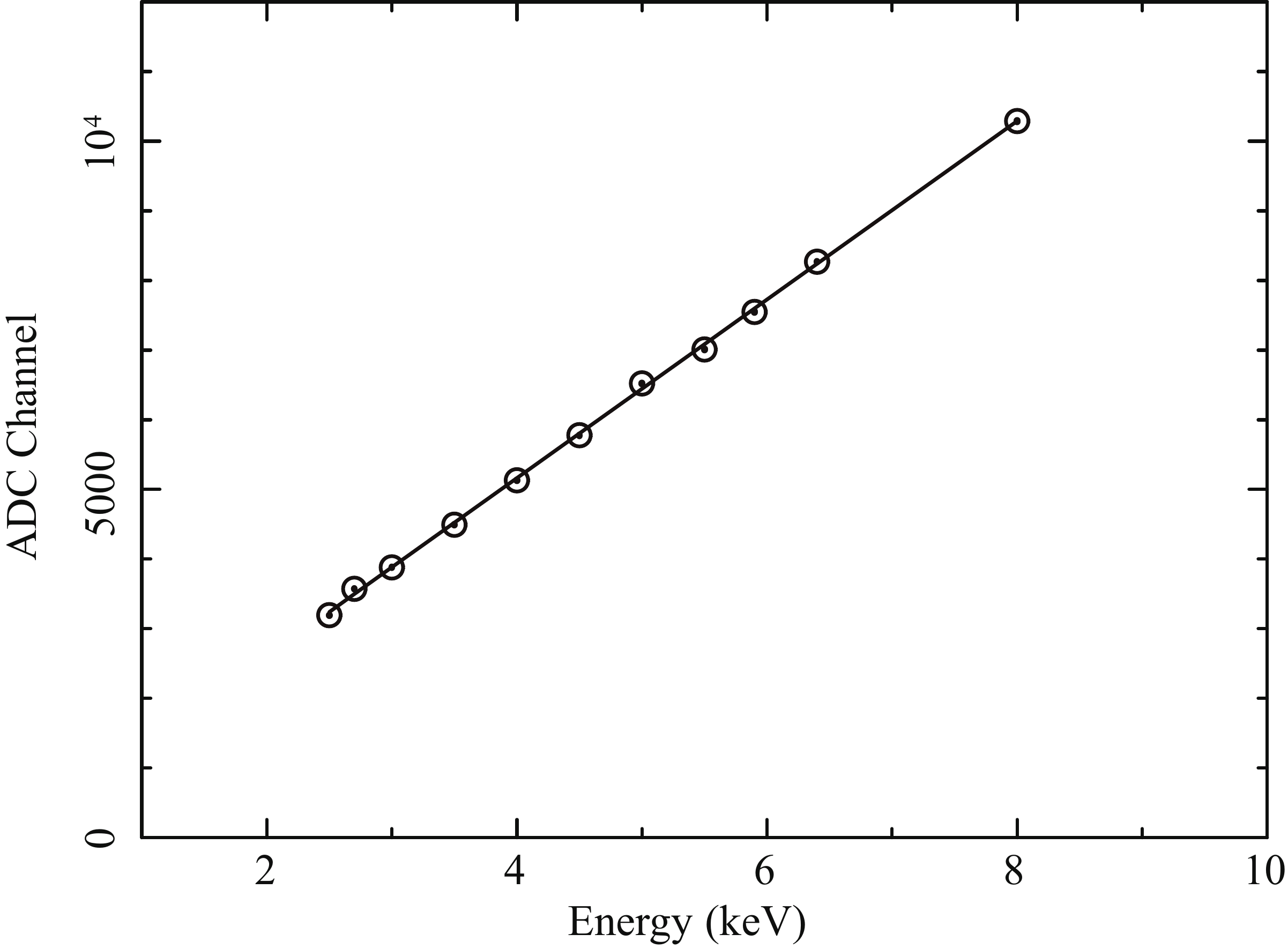}
\caption{The calibration curve of the ADC channel versus input energy obtained at BNL-NSLS X-19A beam line. The error in each data set is less than 0.5\%. The black line is the best-fit linear model; ADC = 36 + 1281 $\times$ $E$ (keV).}
\label{fig4}
\end{center}
\end{figure}

\subsection{Overview of the PRAXyS Polarimeter}
The PRAXyS polarimeter employs a segmented approach, with four
identical readouts arranged in series, parallel to the X-ray beam, as illustrated in Fig.~\ref{fig1}. The
four readouts share a common gas volume as well as a common set of field shaping electrodes \cite{joe2012}.  The
measurements presented in this paper are made with a single readout, identical to the flight design, and a flight-equivalent set of field
shaping electrodes, sized for a single readout.  

A block diagram of the test set-up is shown in Fig.~\ref{fig2}. The active volume for incident X-rays is 78(l) $\times$ 30(w) $\times$
20(h) mm$^3$ between the GEM and the drift electrode. However, the maximum drift distance for these experiments is defined by the size of
the beryllium entrance window, which is 14 mm. The polarimeter was filled with dimethyl ether (DME) to a pressure of 193.5 Torr at a
temperature of 299.4 K. The pitch of the readout strips (121 $\mu$m) and the drift velocity multiplied by the sampling time (50 ns) defines
the dimensions of the image pixels. Magboltz\footnote{http://consult.cern.ch/writeup/magboltz/}
was used to estimate the drift field required to obtain a drift velocity for a two dimensional image with square pixels (A drift
velocity of 0.242 cm~$\mu$s$^{-1}$ requires an applied drift field of 196 V~cm$^{-1}$). The drift velocity calculation was confirmed by measurement. 
A transfer field of 660 V~cm$^{-1}$ was applied across the 0.25 mm transfer gap between the GEM and the readout strips. This value is a compromise
between efficient charge collection and asymmetric diffusion in the transfer gap. Figure \ref{fig3} shows the typical energy spectrum
obtained at 6.4 keV, with full-width-at-half-maximum (FWHM) energy resolution of 16\%. As in the flight design, the signals
from the strip electrodes were read out via an APV25 Application Specific Integrated Circuit (ASIC) \cite{apv25}.  Signal processing outside the gas volume was performed with ground support equipment (GSE) electronics. The GEM cathode signal was amplified and shaped by an ORTEC 142AH preamplifier and an ORTEC 671 shaping amplifier and digitized by a TENNELEC TC451 constant fraction discriminator to generate a trigger signal in the Field Programmable Gate Array (FPGA) to readout out the ASIC. A more detailed description of the readout system is provided in Black
et al. 2010 \cite{kevin2010}. 

\subsection{Beamline measurement}
The response to polarized X-rays was measured at eleven different monochromatic energies from 2.5 to 8.0 keV.  The X-ray beam was collimated to $<$ 0.25 mm, and $\sim$1.5$\times$10$^5$ counts were obtained at each energy except at 2.5 keV, where we collected $\sim$3$\times$10$^4$ counts.  Data were obtained at a range of drift heights. Figure \ref{fig4} shows that the relation between pulse height and energy is quite linear. The polarimeter was inclined at approximately $-$45$^{\circ}$ relative to the polarization vector of the synchrotron beam.

The polarization of the synchrotron beam itself was independently measured using a scattering polarimeter which consists of a cylindrical Be scatterer and perpendicularly placed solid-state detector. The data at 7.8 and 10 keV are consistent with a beam polarization 
of 94\% (Enoto et al. 2014 \cite{enoto2014}, Appendix A). We assume that the beam polarization is 94\% at all energies.

\section{Analysis and Results}
\subsection{Data processing}
An image of each photoelectron track is formed from pixels consisting
of 30 strips $\times$ 30 time bins, corresponding to 3.63 $\times$ 3.63 mm$^2$. Both the offset (pedestal) value for each strip electrode
and the common mode noise (median pulse height response of each strip not contained in the event) is subtracted prior to constructing the
image. Only pixels that contain charge greater than 3 times the root mean square (RMS) noise are included in the image analysis. The
APV25 ASIC applies a 50 ns shaping time to each signal. The response is de-convolved with the measured ASIC response
$h(t)$ to an internal test pulse. If the input $f(t)$ and output signal $o(t)$ of a time series relate to $h(t)$ according to:

\begin{equation}
o(t) = \int h(t - t^\prime)f(t^\prime)dt^\prime .
\end{equation}
And the Fourier transformations $O$, $H$ and $F$ in the Fourier space $T$
are: 
\begin{equation}
O(T) = H(T) \cdot F(T) .
\end{equation}
The original input signal is derived as the inverse-Fourier
transformation product of $O(T)/H(T)$ (deconvolution of electronics
response).

The required transfer field for efficient charge collection results in an asymmetry in the intrinsic transverse and longitudinal diffusion. The sampling of the diffusion further increases the asymmetry, as the transverse diffusion is a significant fraction of a pixel (defined by readout strip pitch) while the longitudinal diffusion is a negligible fraction of a pixel (defined by drift velocity in the transfer gap multiplied by the sampling time). The asymmetric diffusion effects are accounted for by applying a Gaussian convolution in the time axis.

We multiply the Fourier transformation $G$ of the Gaussian $g(t)$ in the Fourier space,

\begin{equation}
\hat{F}(T) = O(T) \cdot G(T) / H(T) .
\end{equation}
Thus, the corrected input signals using a Gaussian convolution $g(t)$ is
the inverse-Fourier transformation of $\hat{F}(T)$.

The standard deviation, $\sigma_t$, of the Gaussian convolution was calibrated at the BNL-NSLS beamline using polarized 2.7 keV X-rays. The resulting polarization angle as a function of convolution sigma is shown in Fig. 5. We performed an iterative process to determine the correct $\sigma_t$. Since the polarimeter was inclined at approximately $-$45$^{\circ}$ relative to the polarization vector of the beam, we initially used a best-fit Gaussian width of 39.4 $\mu$m which corresponds to an angle of $-$45$^{\circ}$. This resulted in an average angle of $-$46.0 over the energy range. We iterated the process, this time using a best-fit Gaussian width of 38.9 $\mu$m which equalized the polarization angle to $-$46.1$^{\circ}$.
%The standard deviation, $\sigma_t$, of the Gaussian convolution was calibrated using a polarized 2.7 keV X-ray beam taken at the BNL-NSLS. Since the polarimeter was inclined at $-$45$^{\circ}$ relative to the polarization vector of the beam, we searched for the the best $\sigma_t$ which equalizes the estimated polarization angle to $-$45$^{\circ}$. The results are shown in Fig.~\ref{fig5}. The best-fit Gaussian width is 39.4 $\mu$m, which corresponds to 1/3 of a pixel.

\begin{figure}[!t]
\begin{center}
\includegraphics[clip,width=6.5cm,angle=0]{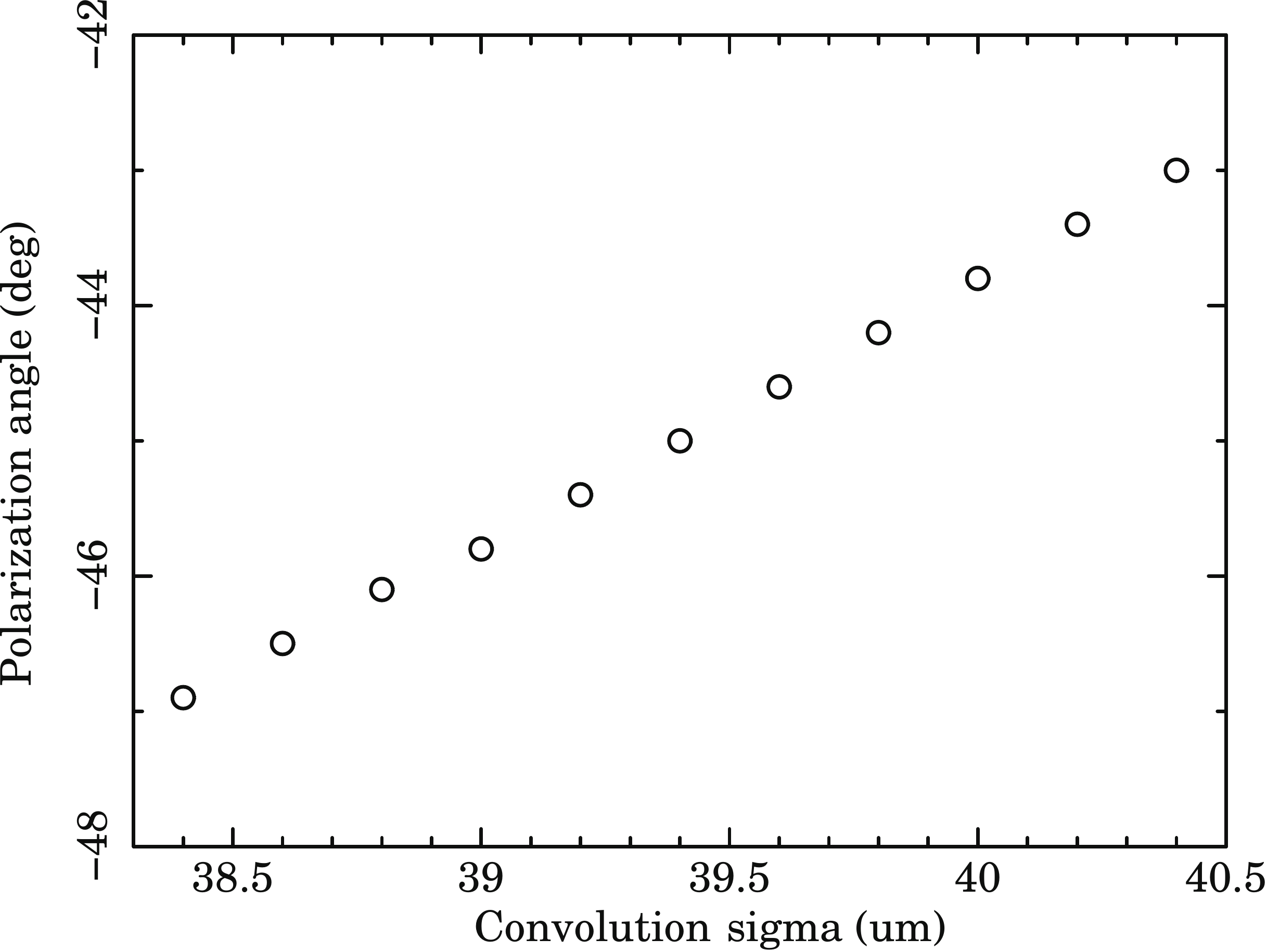}
\caption{Polarization angle of a 2.7 keV polarized X-ray beam as a function of the width of the Gaussian convolution along the time axis. 
}
\label{fig5}
\end{center}
\end{figure}

\begin{figure}[!t]
\begin{center}
\includegraphics[clip,width=9.5cm,angle=0]{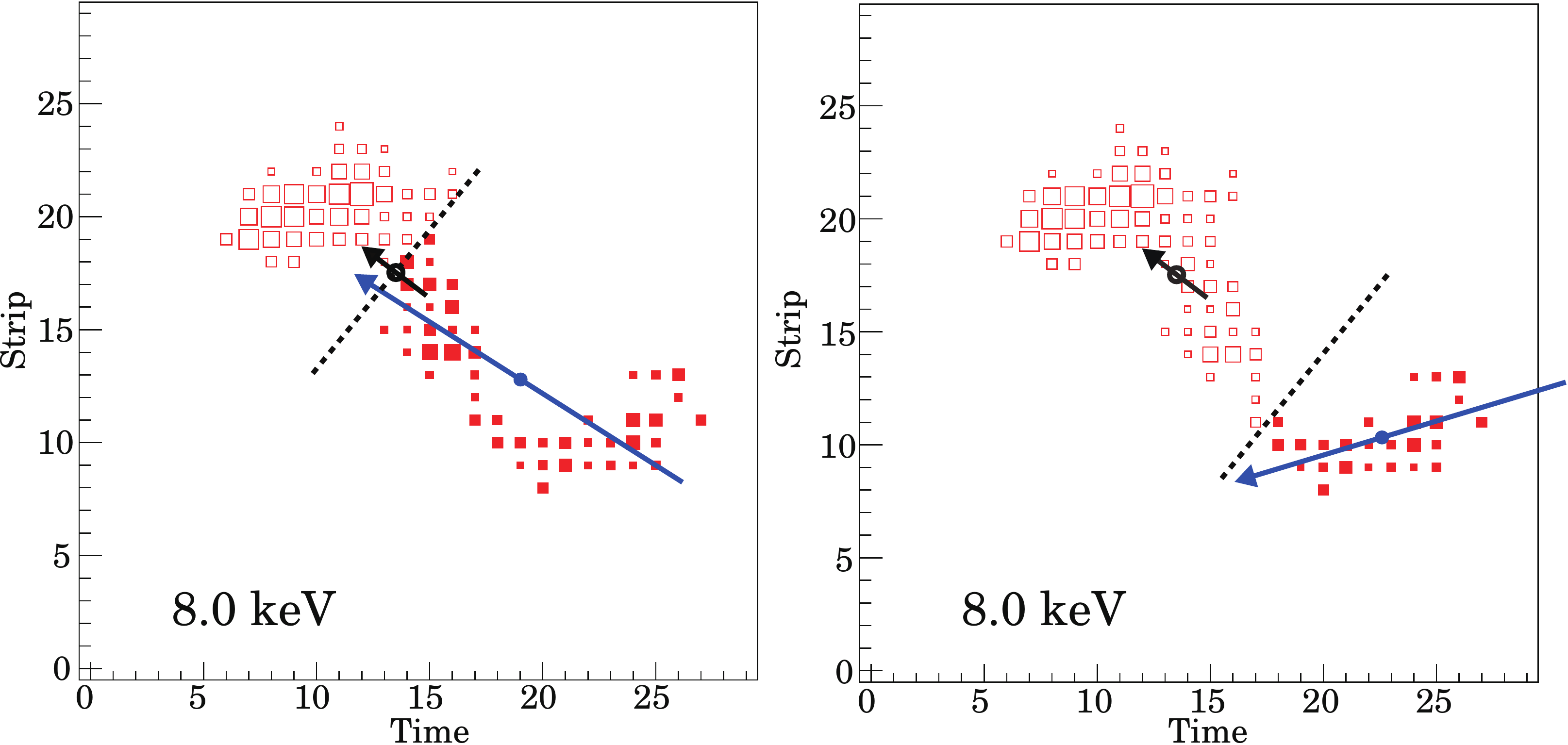}
\caption{Previous track reconstruction algorithms \cite{kevin2007} (left panel) use the initial half of the track. The track is separated along the minor axis of the charge distribution (dashed line) that is perpendicular to the major axis (black arrow) and passed through the centroid of the charge distribution (black dot). We employ a new method that iteratively cuts the charge distribution until a variance/skewness test that identifies straight track segments is satisfied. Only the filled pixels in the right panel are used in the final estimate. For this track, the estimated direction shifts by $\sim$50 degrees between the standard two stage reconstruction and our improved method.
}

\label{fig6}
\end{center}
\end{figure}

\begin{figure}[!t]
\begin{center}
\includegraphics[clip,width=8.0cm]{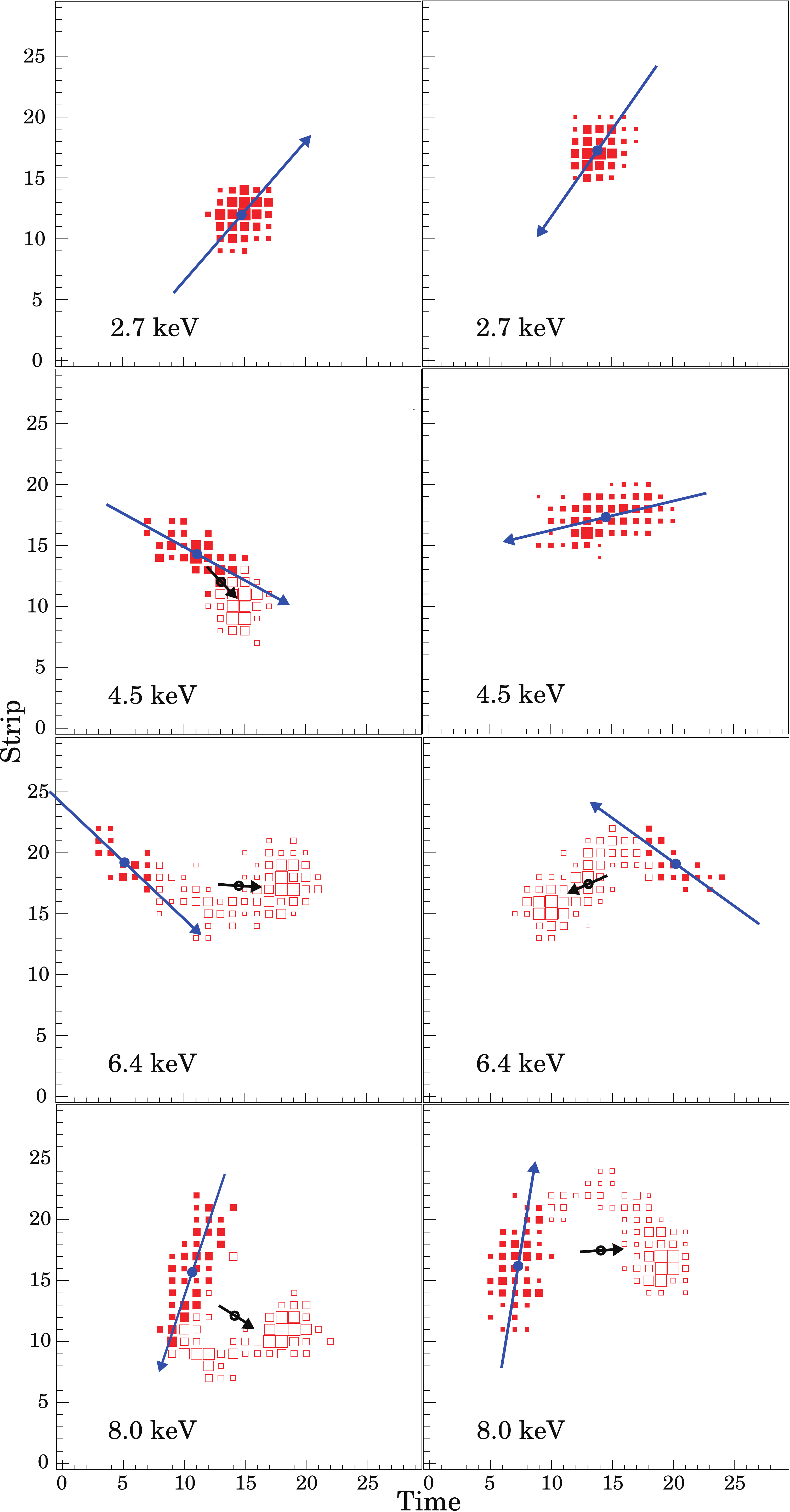}
\caption{Same as the right panel of the Fig.\ref{fig6} but for four energies.
}
\label{fig7}
\end{center}
\end{figure}

\subsection{Angular reconstruction}
Previous analysis has used a two-stage moments analysis to estimate photoelectron directions for individual events \cite{kevin2007}.   In the first stage, the principal axes are estimated using the second order central moments of the charge distribution from the entire track. The previous method is illustrated in the left panel of Fig.~\ref{fig6},  where the size of each symbol is proportional to the recorded charge, and each symbol corresponds to a single 121 $\mu$m resolution element. The black arrow Fig.~\ref{fig6} (left) shows the direction of the major axis; the dot represents the centroid of the charge distribution. For tracks with high eccentricity, the track is divided along the minor axis (dashed line) and the half with higher charge density (the Bragg peak, which represents the end of the track) is ignored. The photoelectron direction is then estimated from moments fit to the first half of the track (blue arrow). However, for long and curved tracks as in Fig.~\ref{fig6}, the two-stage estimate may still be inaccurate.  We have developed a new image reconstruction method \cite{kitaguchi2016} summarized here.

First, the second moment (variance) and the third moment (skewness) along the major and minor principal axes are used to judge whether or not a track is curved. For curved tracks, the charge distribution is repeatedly cut off in 0.5 pixel steps along the major axis of the entire charge distribution until the variance and skewness of the remaining portion of the image are below set thresholds (Fig.~\ref{fig6} right). Lastly, the initial angle of the photoelectron track is reconstructed from the central moments method using the initial part of the track that satisfied the variance/skewness conditions.
The blue arrow in the right panel of Fig.~\ref{fig6} shows the result obtained by the new method, which is noticeably more accurate than the two-stage reconstruction in the left panel. The improved estimate of the track direction leads to higher values of $\mu$.

\subsection{Event selection}

In principle, the reconstructed photoelectron emission angle gives the polarization state of every X-ray.  However, instrumental effects can obscure the emission direction, especially at lower energies, eliminating the polarization information for some X-rays.  These events essentially form an unpolarized background.  The obscuring effects include Coulomb scattering, electron diffusion, charge associated with an isotropically emitted Auger electron and foreshortening of tracks when projected onto the readout plane.  These instrumental effects are confirmed by Monte-Carlo simulation \cite{kitaguchi2014} and are characterized by low eccentricity.

To minimize the MDP, we maximize $\mu \sqrt{\epsilon}$, which is the figure of merit explained in Section 2, by excluding events with eccentricity below an experimentally determined and pulse height dependent threshold. The event threshold, $e_\text{th}$, for the measured pulse height, PH, is given by:

%\mathindent=0zw
\begin{eqnarray}
%\[
  e_{\text{th}} = \left\{ \begin{array}{ll}
    {\scriptstyle 0.48} & {\scriptstyle(\text{PH} < 5 ~\text{keV})} \\
    {\scriptstyle -1.1604 + 0.4882 \cdot E - 0.0321 \cdot E^2 }& {\scriptstyle(5 ~\text{keV} \leq \text{PH} \leq 8 ~\text{keV})}\\
    {\scriptstyle 0.69} & {\scriptstyle(\text{PH} > 8 ~\text{keV}).}
  \end{array} \right.
%\]
\end{eqnarray}
%\mathindent=1zw
The fractional increase in $\mu$ is larger than the fractional decrease in $\sqrt{\epsilon}$, thus improving the overall figure-of-merit $\mu\sqrt{\epsilon}$.
Similar measurements for an alternate detector geometry \cite{li2015, Weisskopf2016} also demonstrate the benefit of an eccentricity based selection.

\begin{figure}[!t]
\begin{center}
\includegraphics[clip,width=8.0cm]{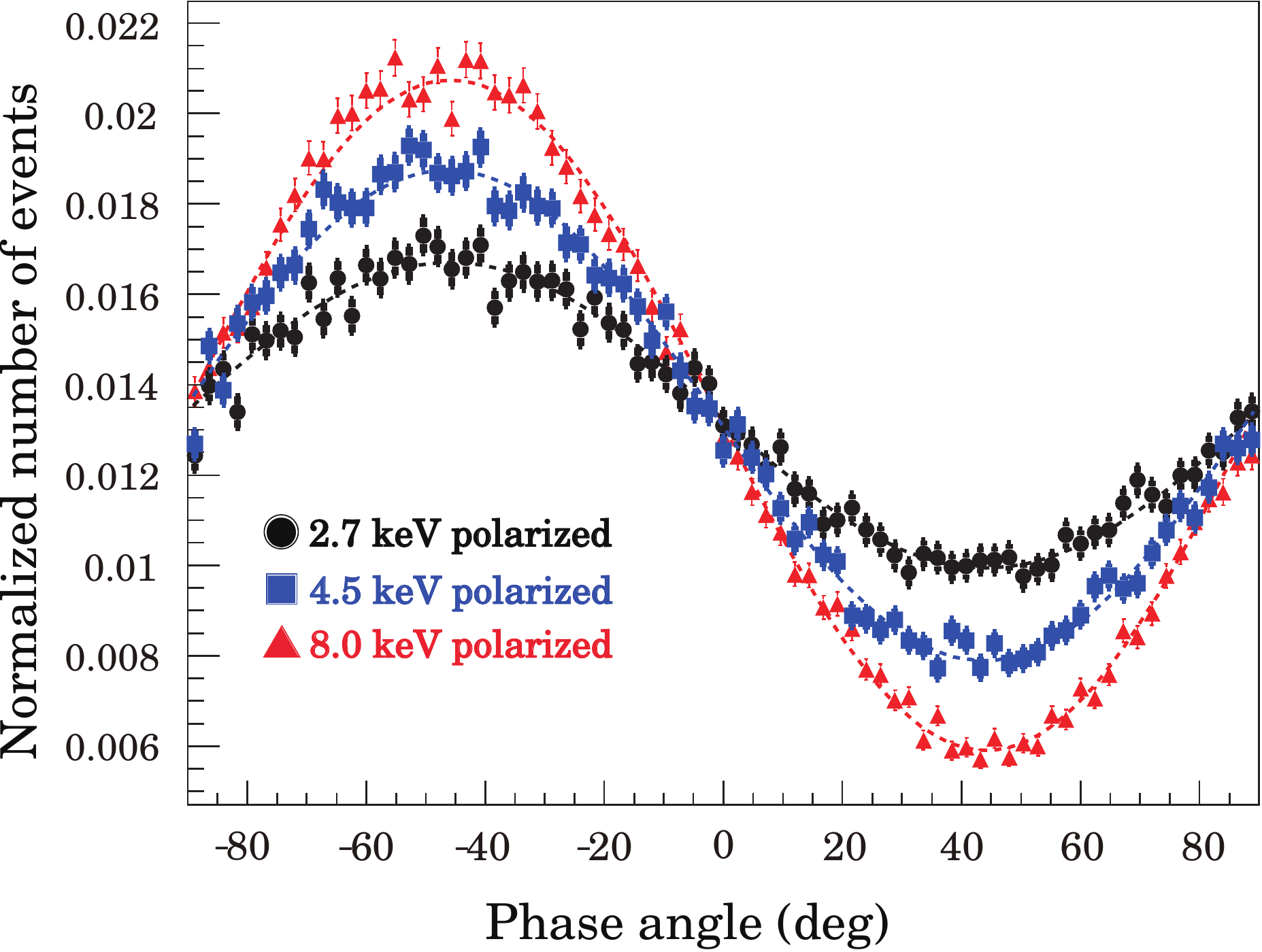}
\caption{Modulation curves for three energies at an average drift height of 8 mm(averaged from 6 mm, 8 mm and 10 mm). The effective modulation factors $\mu$ of 2.7 keV polarized (black), 4.5 keV polarized (blue) and 8.0 keV polarized (red) are 26.92$\pm$0.66\%, 43.38$\pm$0.59\% and 59.14$\pm$0.55\%, respectively.}
\label{fig8}
\end{center}
\end{figure}

\begin{table*}[t]
\begin{center}
\caption{\label{tab1} Best-fit parameters for the modulation curves. All errors denote 90\% error level.}
\begin{tabular}{ccccc}
\hline
\hline
Energy & $\mu$ & $\phi_{0}$ & $f^{\ast}$ & $\chi^2$/d.o.f$^\dagger$\\
(keV) & (\%) & (degree) & & \\
\hline
2.5 & $24.53\pm2.10$ & $-45.10\pm2.67$ &0.76 &19.70/17 \\
2.7 & $26.92\pm0.66$ & $-45.90\pm0.76$ &0.76 &94.85/97 \\
3.0 & $32.61\pm0.65$ & $-47.26\pm0.62$ &0.79 &105.65/97 \\
3.5 & $36.63\pm0.61$ & $-45.82\pm0.53$ &0.85 &89.78/97 \\
4.0 & $40.90\pm0.60$ & $-45.85\pm0.46$ &0.91 &125.45/97 \\
4.5 & $43.38\pm0.59$ & $-46.35\pm0.43$ &0.92 &132.08/97 \\
5.0 & $46.00\pm0.58$ & $-46.47\pm0.40$ &0.91 &142.29/97 \\
5.5 & $49.24\pm0.58$ & $-45.88\pm0.38$ &0.90 &122.61/97 \\
5.9 & $52.48\pm0.57$ & $-46.09\pm0.35$ &0.91 &141.10/97 \\
6.4 & $54.42\pm0.57$ & $-46.44\pm0.34$ &0.89 &127.62/97 \\
8.0 & $59.14\pm0.55$ & $-45.77\pm0.31$ &0.87 &112.24/97 \\
\hline
\multicolumn{5}{l}{$\ast$ : the signal acceptance after eccentricity cuts.} \\
\multicolumn{5}{l}{$\dagger$ : degrees of freedom.} \\
\end{tabular}
\end{center}
\end{table*}

\begin{figure*}[!t]
\begin{center}
\includegraphics[clip,width=16.0cm]{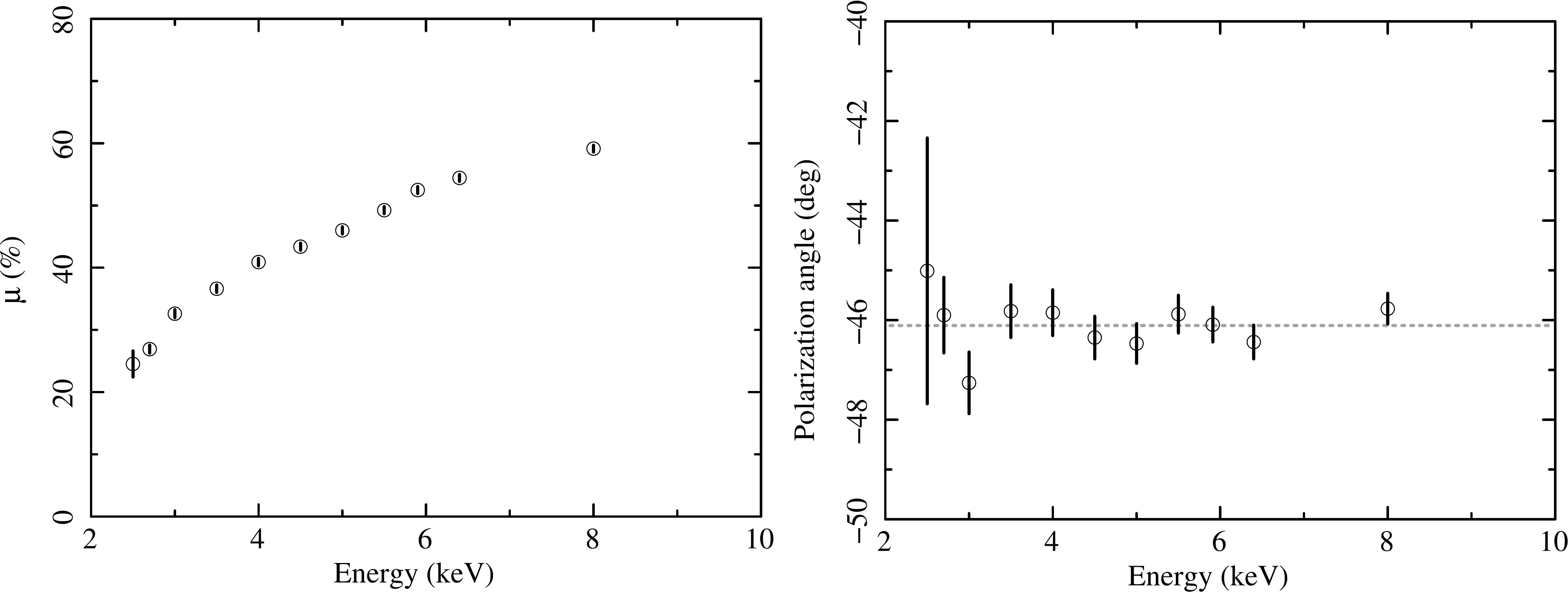}
\label{fig9}
\caption{(Left) Modulation factor as a function of an incident X-ray energy. (Right) Same as left panel but for polarization angle. Dotted line shows the mean polarization angle.}
\end{center}
\end{figure*}

\begin{figure*}[t]
\begin{center}
\includegraphics[clip,width=16.0cm]{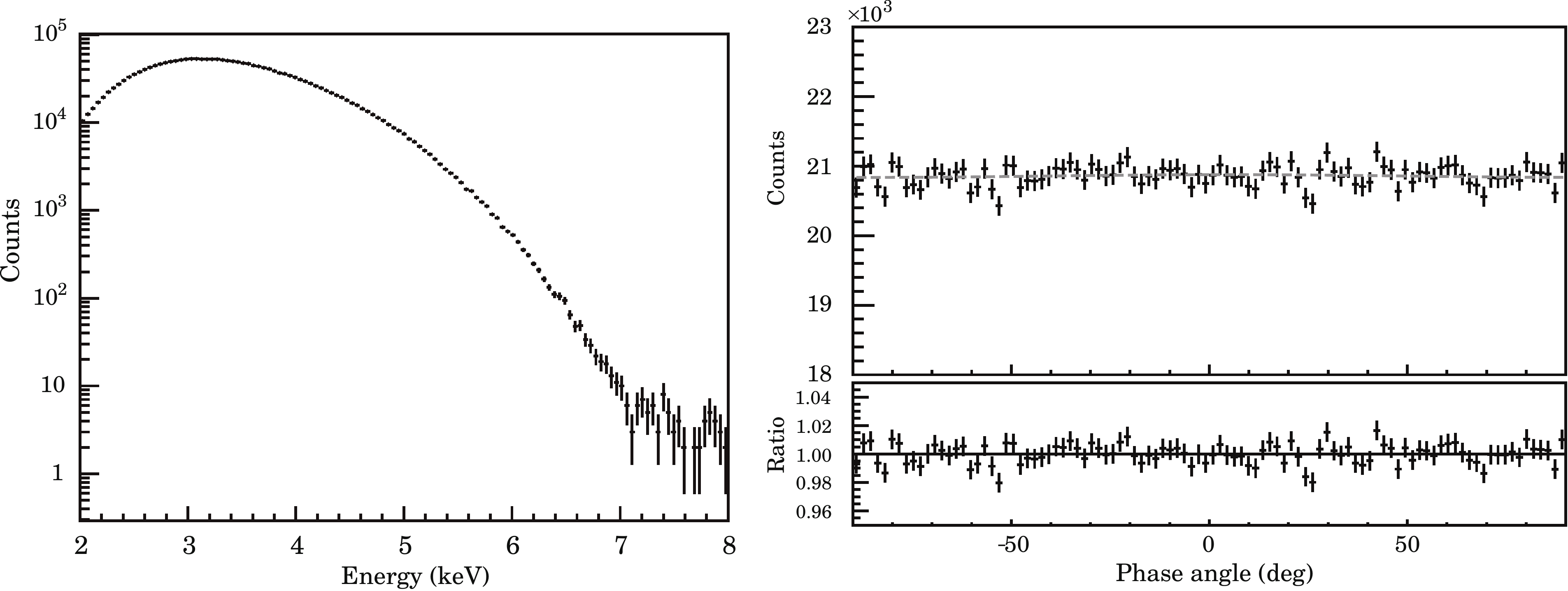}
\label{fig8}
\caption{Spectrum (left) and modulation curve (right) for the unpolarized broadband Bremsstrahlung source. The modulation factor $\mu$ = 0.10 $\pm$ 0.16\% with $\chi^2$ / d.o.f = 113.52 / 97.}
\end{center}
\end{figure*}

\subsection{Performance tests with polarized X-rays at BNL-NSLS}
Examples of track images taken at the BNL-NSLS, after the data
processing described in the previous section, are shown in
Fig.~\ref{fig7}. Figure 8 shows histograms of emission angles for three different polarized X-ray energies. 
The fitting results show that the measured effective modulation factors, $\mu$, are 26.92\%, 43.38\% and 59.14\% at 2.7, 4.5 and 8.0 keV, respectively. We show the summary of $\mu$ and polarization angle as a function of energy in
Fig.~9 and Table \ref{tab1}. The signal acceptance after eccentricity cuts is also shown in Table \ref{tab1}. The measured values of $\mu$ versus $E$ are quite similar to those of Li et al. 2015 \cite{li2015}. This is not surprising as our geometry and Li et al. have similar products of pressure and pixel size, so that track lengths, measured in pixels, is similar. 
Figure 9 (right) shows that the reconstructed polarization angle is independent of energy. The mean polarization angle $\phi$ is
$-$46.1$^{\circ}$ $\pm$ 0.1$^{\circ}$, it is consistent with the expected angle within statistical error. 

\subsection{Performance tests with unpolarized X-ray}
With modulation similar to a pixel polarimeter, the greater quantum efficiency of the TPC polarimeter will allow it to achieve higher
sensitivity only if systematic errors that create false modulation are small compared to the statistical limits.  PRAXyS employs multiple strategies,
including instrument rotation, to eliminate such errors.   

We collected 2.6 million events from an unpolarized broadband Bremsstrahlung spectrum that peaks around 3 keV with a 5 keV endpoint
shown in the left panel of Fig.~10. For the PRAXyS mission design, the worst case pointing error (including alignment terms) is 1 arcmin,
which corresponds to 1.3 mm at the center of the detector. Therefore, to simulate the rotation, we took 36 measurements, each with 64,000 events
which went around the compass in 10 degree steps.  The mean drift distance, $d$, from the interaction point to the GEM, is $d = 8 +
1.3\sin\theta ~(\text{mm})$. An 8 mm drift height corresponds to the optical axis of the detector. If the detector were rotating about a different axis, the apparent mean drift distance would vary as above. We simulate this rotation by moving a collimated pencil beam in a circle on the detector aperture. Theta is also the amount by which each data set must be shifted to transform detector coordinates to laboratory (or sky) coordinates. We co-added the data in effective sky coordinates using $\theta$ as the ephemeris.

We reconstructed the data following the analysis steps described in section 3.1, 3.2 and 3.3. After the eccentricity cut based on pulse
height, 2.1 million events remained. The measured modulation factor is 0.10\% $\pm$ 0.16\% (90\% confidence level) shown in Fig.~10
(right). The MDP associated with these data, using an average value of $\mu$ weighted by the counts spectrum in Fig.~10 (left) is 0.87\%.
Thus the polarimeter is capable of making statistics limited polarization detections at levels below 1\%. For true polarization fractions of 2\% (5\%), the polarimeter will make 4.6 $\sigma$ (11.6 $\sigma$) detections from similar datasets ($2.1 \times 10^6$ counts after eccentricity selection).

\section{Conclusion}
We evaluated the performance of the prototype polarimeter for PRAXyS using the linearly-polarized X-ray source at BNL-NSLS between 2.5 and 8.0 keV. With unpolarized X-rays, we measured an upper limit to the expected systematic errors that would lead to false polarization. These measurements demonstrate that the polarimeter meets or exceeds the sensitivity required for PRAXyS to reach its scientific goals. The results are summarized below:

\begin{itemize}
  \item After the image reconstruction of photoelectron track and the optimized eccentricity cut, the modulation factors, $\mu$, of the PRAXyS polarimeter are 27\%, 43\% and 59\% at 2.7, 4.5 and 8.0 keV, respectively. 
 
  \item Measured polarization angles are constant relative to incident energy. For small polarization fractions, the error on the polarization angle will be limited by statistics. For large polarization fractions and significant measurements, the maximum error will be less than 1$^{\circ}$. These values exceed the requirements levied on the PRAXyS polarimeters.

  \item False modulation is not detected in a continuum dominated spectrum, representative of that expected from astronomical observations, with over 2 million counts, which is comparable to the number of photons needed to detect a 1\% polarization.

\end{itemize}

\section*{Acknowledgments}
This work was partially supported via proposal 13-APRA13-0141 in response to 
the NASA  Astrophysics Research and Analysis solicitation NNH13ZDA001N-APRA and
MEXT KAKENHI Grant Number 24105007. The authors would like to acknowledge the support of Syed Khalid at the X19A beamline at the BNL-NSLS. W.B. Iwakiri was supported by JSPS KAKENHI, Grant-in-Aid for JSPS Fellows, 25-5312. We also thank the two anonymous referees whose comments have helped us improve the presentation of these results.
\bibliography{ref.bib}

\end{document}